# Role of further-neighbor interactions in modulating the critical behavior of the Ising model with frustration

R. M. Liu[1], W. Z. Zhuo[1], S. Dong[2], X. B. Lu[1], X. S. Gao[1], M. H. Qin[1 *], and J. -M. Liu[1,3 *]

[1]*Institute for Advanced Materials and Guangdong Provincial Key Laboratory of Quantum Engineering and Quantum Materials, South China Normal University, Guangzhou 510006, China*

[2]*Department of Physics, Southeast University, Nanjing 211189, China*

[3]*Laboratory of Solid State Microstructures, Nanjing University, Nanjing 210093, China*

[**Abstract**] In this work, we investigate the phase transitions and critical behaviors of the frustrated $J_1$-$J_2$-$J_3$ Ising model on the square lattice using Monte Carlo simulations, and particular attention goes to the effect of the second next nearest neighbor interaction $J_3$ on the phase transition from a disordered state to the single stripe antiferromagnetic state. A continuous Ashkin-Teller-like transition behavior in a certain range of $J_3$ is identified, while the 4-state Potts-critical end point $[J_3/J_1]_C$ is estimated based on the analytic method reported in earlier work [Jin *et al*., Phys. Rev. Lett. 108, 045702 (2012)]. It is suggested that the interaction $J_3$ can tune the transition temperature and in turn modulate the critical behaviors of the frustrated model. Furthermore, it is revealed that an antiferromagnetic $J_3$ can stabilize the staggered dimer state via a phase transition of strong first-order character.

---

*Electronic mail: qinmh@scnu.edu.cn (M. H.) and liujm@nju.edu.cn (J. M.)

## I. Introduction

Symmetry breaking at a thermal phase transition decides orders. For example, the spontaneous breaking of $Z_2$-symmetry in the well-known frustrated two-dimensional (2D) antiferromagnetic (AFM) $J_1$-$J_2$ model on the square lattice leads to the Néel state (Ising AFM state, Fig. 1(a)) for $J_2/J_1 < 1/2$.[1-4] The phase transition between the Ising AFM state and a disordered state belongs to the 2D Ising universality class. Furthermore, for $J_2/J_1 > 1/2$, a $Z_4$-symmetry can be broken below the critical temperature ($T \leq T_C$), resulting in the single stripe AFM state (Fig. 1(b)).[5] More interestingly, the nature of this phase transition cannot be directly obtained from the symmetry of the order parameter, and different phase-transition scenarios apply depending on the value of $J_2/J_1$, as will be introduced below.[6]

In the 1980s, it was generally believed that the transition to the single stripe order is continuous but with varying critical exponents for different $J_2/J_1$.[7-10] However, this point of view was suspected by the mean-field calculation in 1993 which found a first-order transition for a certain region of $J_2/J_1$.[11] Subsequently, the existence of the first-order transition was further confirmed by Monte Carlo (MC) simulations.[12-14] In particular, the transition for $1/2 < J_2/J_1 \leq 1$ was suggested to be of weak first order based on a double-peak structure in energy histograms. Furthermore, it was suggested that the transition for large $J_2/J_1 > 1$ is continuous with the Ashkin-Teller (AT) criticality.[14,15]

More recently, the nature of the stripe phase transition in the $J_1$-$J_2$ model was clearly elucidated by employing a combination of MC simulations and analytical methods, and the transition point between the two scenarios (first-order and AT-like transitions) was estimated to be at a value of $J_2/|J_1| \approx 0.67$.[6,16] The phase transition at this critical point is in the 4-state Potts universality class, and those for $1/2 < J_2/J_1 < 0.67$ are of weak first order.[17] More interestingly, a pseudo-first-order behavior was uncovered for $0.67 \leq J_2/|J_1| \leq 1$, similar to that of the AT model in certain parameter region. It was demonstrated that some signatures were not sufficient as proofs of first-order transition, leading to the former overestimation of the region of first-order character in the $J_1$-$J_2$ model. This behavior was also verified by the large-scale MC simulations which show that the first-order signals in the energy histograms vanish at large system sizes $L \sim 2000$ at $J_2/J_1 = 0.8$.[18]

While the phase transitions and critical behaviors of the frustrated $J_1$-$J_2$ model are being progressively uncovered (as summarized in Fig. 1(c)), researches proceed in the models with further neighboring interactions. The study of the role of further neighboring interactions on the critical behaviors becomes important from the following two viewpoints. On one hand, the single stripe order and various transition behaviors have been experimentally reported in most of the iron-based superconductors. The exchange interaction paths in relevant materials are very complicated. Further neighboring interactions may be available and play an important role in determining the magnetic properties due to the spin frustration.[19] For example, a nonzero coupling $J_3$ between the third nearest neighbors is suggested to be important for the magnetic properties in iron chalcogenides such as FeTe.[20] On the other hand, this study can also contribute to the development of statistical mechanics and solid state physics. For example, more interesting AFM orders such as staggered dimer and double stripe states (shown in Fig. 2(a) and 2(b), respectively) can be stabilized by an AFM $J_3$, and the transition behaviors are also very attractive. Furthermore, the interesting scenarios for the single stripe phase transition in the $J_1$-$J_2$ model discussed above promote the development of phase transition theory. However, there are still open questions such as whether these scenarios hold true for the model with further neighboring interactions, and how the critical behaviors are determined. In some extent, these questions are related to the universality of the phase transition, and definitely deserve to be checked in details.

In this work, we study the frustrated $J_1$-$J_2$-$J_3$ Ising model on the 2D square lattice to unveil the role of $J_3$ in modulating the critical behaviors. The MC simulated results show that the critical exponents of the single stripe AFM transition vary with the increasing magnitude of ferromagnetic (FM) $J_3$ (at $J_2/J_1 = 0.8$, for example). Similarly, a pseudo-first-order behavior is observed for small AFM $J_3$, and the 4-state Potts critical end point is also reasonably estimated. Furthermore, the transition from a disorder state to the staggered dimer state is investigated, which exhibits a strong first-order behavior.

The rest of this paper is organized as follows. In Sec. II the model and the simulation method will be presented and described. Section III is attributed to the simulation results and discussion. The conclusion is presented in Sec. IV.

## II. Model and method

The model Hamiltonian can be written as:

$$H = J_1 \sum_{\langle ij \rangle_1} S_i S_j + J_2 \sum_{\langle ij \rangle_2} S_i S_j + J_3 \sum_{\langle ij \rangle_3} S_i S_j, \tag{1}$$

where $J_1 = 1$ is the unit of energy, $S_i = \pm 1$ is the Ising spin with unit length on site $i$, $\langle ij \rangle_n$ denotes the summations over all the $n$-th nearest neighbors with coupling $J_n$.

For a description of the single stripe order, the order parameter $m_s$ can be defined as:

$$m_s^2 = m_x^2 + m_y^2, \tag{2}$$

with

$$m_{x,y} = \frac{1}{N} \sum_i (-1)^{i_{x,y}} S_i, \tag{3}$$

where $(i_x, i_y)$ are the coordinates of site $i$ on an $N = L \times L$ ($24 \leq L \leq 256$) periodic lattice. In order to understand the nature of the phase transition, we calculate the Binder cumulant $U_s$:

$$U_s = 2\left(1 - \frac{1}{2} \frac{\langle m_s^4 \rangle}{\langle m_s^2 \rangle^2}\right), \tag{4}$$

and susceptibility $\chi_s$:

$$\chi_s = N\left(\langle m_s^2 \rangle - \langle |m_s| \rangle^2\right)/T, \tag{5}$$

where $\langle \ldots \rangle$ is the ensemble average.

The staggered dimer state is eight-fold degenerate, and the order parameter $m_d$ can be similarly defined as:

$$m_d^2 = \sum_{k=1}^{4} m_k^2, \tag{6}$$

with

$$m_k = \sum_i A_k(i_x, i_y) S_i, \tag{7}$$

where the value of $A_k(i_x, i_y)$ depends on the coordinates of site $i$ and the spin configurations of the ground states. Take the configuration shown in Fig. 2(a) as an example, $A(i_x, i_y) = 1$ for up-spins, and $A_k(i_x, i_y) = -1$ for down-spins. Furthermore, the Binder cumulant $U_d$ is also calculated.

Our simulation is performed using the standard Metropolis algorithm and the parallel tempering algorithm.[21,22] We take an exchange sampling after every 10 standard MC steps.

The initial $5\times10^5$ MC steps are discarded for equilibrium consideration and another $5\times10^5$ MC steps are retained for statistic averaging of the simulation. Generally, we choose $J_2 = 0.8$ and change $J_3$ in computation in studying the single stripe phase transition. It is noted that $J_3$ has no effect on the competition between the stripe state and the Ising AFM state,[23] and the former state occupies the whole studied $J_3$ region ($J_3 < 0.15$). Furthermore, it will be checked later that the choice of $J_2$ never affects our conclusion. On the other hand, $J_2 = 0.5$ is selected to study the phase transition to the staggered dimer state.

### III. Simulation results and discussion

*A. Varying critical exponents with ferromagnetic $J_3$*

First, we study the effect of FM $J_3$ on the phase transition and its critical behaviors. Figs. 3(a) and 3(b) show the simulated $U_s$ as a function of $T$ at $J_3 = -0.2$ and $-0.5$ for different $L$. For continuous phase transitions, the Binder cumulant for different $L$ usually crosses at the critical point. From the well common defined crossing points, we estimate $T_C = 2.218(5)$ at $J_3 = -0.2$ and $T_C = 3.131(5)$ at $J_3 = -0.5$. In fact, it is rather clear that the single stripe order can be further favored by a FM $J_3$, and the transition point shifts toward high $T$ when the magnitude of $J_3$ is increased, as shown in our simulations.

In the AT model, the critical exponents $v$ and $\gamma$ vary with the magnitude of the frustration, while the ratio of $\gamma/v = 7/4$ keeps constant.[24] This critical behavior is also observed in our simulations of the $J_1$-$J_2$-$J_3$ Ising model, as shown in Fig. 4. The critical exponents are estimated based on the standard finite-size scaling fact that the slope of $U$ vs $T$ at $T_C$, $dU/dT(T_C)$, is proportional to $L^{1/v}$, and $\chi_{max}$ is proportional to $L^{\gamma/v}$. It is clearly shown that $v$ increases with the increasing magnitude of $J_3$ and/or $T_C$. For example, $v = 0.83(3)$ at $J_3 = -0.2$ and $v = 0.88(5)$ at $J_3 = -0.5$ are estimated, as shown in Fig. 4(a). Furthermore, the roughly constant $\gamma/v = 7/4$ is obtained for every FM $J_3$ (Fig. 4(b)), within the limits of acceptable error (at most, ~ 1.8%), demonstrating an AT-like behavior. In Fig. 5, we plot the simulated $U_s$ and $\chi_s$ in the scaling form: $U_s = f(tL^{1/v})$, and $\chi_s = L^{\gamma/v}g(tL^{1/v})$, with $t = (T - T_C)/T_C$, at $J_3 = -0.2$ and $-0.5$. It is confirmed that the single stripe phase transition at $T_C = 2.218(5)$ for $J_3 = -0.2$ is with the critical exponents $v = 0.83(3)$ and $\gamma = 1.45(8)$, and that at $T_C = 3.131(5)$ for $J_3 = -0.5$ is with $v = 0.88(5)$ and $\gamma = 1.54(9)$.

*B. Location of the Potts-critical end point*

If the continuous single stripe phase transition in the $J_1$-$J_2$-$J_3$ model can be mapped to the critical line of the AT model, a 4-state Potts-critical end point is expected at an AFM $J_3$ which destabilizes this order and diminishes $T_C$.[25] In fact, pseudo/weak-first-order behavior is also observed at small AFM $J_3$ for intermediate $L$. Fig. 6 gives the $U_s$ as a function of $T$ for various $L$ at $J_3 = 0.05$ and $0.15$. A negative peak is developed for $L = 128$ at $J_3 = 0.05$ and grows as $L$ further increases, indicating a pseudo/weak-first-order transition behavior, as clearly shown in Fig. 6(a).[26] Furthermore, the system size needed to stabilize a negative peak is decreased when $J_3$ is increased (for example, $L = 24$ at $J_3 = 0.15$, as shown in Fig. 6(b)), demonstrating an enhancing discontinuity of $m_s$.

It is noted that the negative cumulant peak is not a sufficient proof for the first order transition because such a peak can appear also for continuous transitions in spin models such as the 4-state Potts and AT models.[6] However, the Binder crossing value $U^*$ is normally universal and may characterize the universality class of the phase transition. Thus, following earlier works, the 4-state Potts-critical end point of the AT line in the $J_1$-$J_2$-$J_3$ model can be reasonably estimated based on the analysis of the universality of the Binder cumulants ($U^* \approx 0.79$ for the 4-state Potts model). Fig. 7(a) shows the Binder cumulant crossing points for pairs $(L, 2L)$ and $L = \infty$ extrapolated $U_s^*$ for various $J_3$. It is clearly shown that $U_s^*$ decreases with increasing $J_3$. Finally, critical $[J_3/J_1]_C = 0.11 \pm 0.01$ for $J_2/J_1 = 0.8$ is obtained by comparing $U_s^*(J_3)$ with $U^*$ for the 4-state Potts model, as shown in Fig. 7(b).

*C. Single stripe phase transition behaviors and discussion*

The simulated results of the single stripe phase transition for $J_2 = 0.8$ are summarized in Fig. 8(a). The critical temperature for $J_3 \geq 0.05$ is estimated from the position of the peak of $\chi(T)$ curve for the largest $L$.[27] The phase diagram exhibits two regions with different transition behaviors, similar to that of the $J_1$-$J_2$ model. In detail, an AT-like behavior is observed for $J_3 \leq 0.11$, in which $\nu$ and $\gamma$ increase continuously from those of the 4-state Potts model to those of the 2D Ising model, respectively. More interestingly, our simulations also show a close dependence of the values of $\nu$ and $\gamma$ on the transition point $T_C$ for a fixed $J_1$, indicating that this phenomenon may be universal in the single stripe phase transitions in different models. In fact, similar behavior has been observed for other values of $J_2$. Specifically, Fig. 8(b) gives the estimated Potts-critical end points $[J_3/J_1]_C$ for different $J_2$. $[J_3/J_1]_C$ shifts toward high-$J_3$ side as $J_2$ increases, while the transition point $T_C$ at $[J_3/J_1]_C$ is less affected. Furthermore, it is worth noting that the model has a symmetry $(J_1, J_2, J_3) \rightarrow (-J_1, J_2, J_3)$, and a FM $J_1$ would never affect our conclusion.

On one hand, the present work shows that the single stripe phase transition behavior is also dependent on the exchange couplings between distant neighboring spins. The phase transition scenarios uncovered in the $J_1$-$J_2$ Ising model still hold true when $J_3$ interaction is taken into account, further supporting the universality of these scenarios. In addition, it is suggested that the critical behavior may be likely dependent on the transition point which can be detailed modulated through various methods. Of cause, additional proofs should be needed to double-check the universality of the transition pictures in some other frustrated spin models. On the other hand, the single stripe order as the ground state of most of iron-based superconductors has drawn extensive attentions in the past a few years.[28,29] For example, the AFM phase transition in La-O-Fe-As is of first order, while that in BaFe$_2$As$_2$ is continuous.[30] In some extent, the transition scenarios uncovered in the frustrated Ising model on the square lattice may provide useful information in understanding the phase transitions in these materials, although some other degrees of freedom should be taken into account.

*D. Other orders in the phase diagram of the $J_1$-$J_2$-$J_3$ model*

One may note that some other orders can be stabilized by AFM $J_3$ interaction.[31,32] In detail, Fig. 9 gives the ground state phase diagram of the $J_1$-$J_2$-$J_3$ model which can be easily obtained by mean-field method. It is clearly shown that the staggered dimer state is stabilized for AFM $J_3 < 0.5$ (yellow region), and the double stripe state or plaquette state occupy the $J_3 > 0.5$ region. Different with the four-fold degenerated single stripe state, the staggered dimer state is eight-fold degenerate. Thus, the phase transition to the staggered dimer state is expected to be of first order, similar to the transition in the 8-state Potts model. This viewpoint has been confirmed in our simulations. Fig. 10(a) shows the calculated $U_d$ as a function of $T$ for various $L$ at $J_3 = 0.1$ and $J_2 = 0.5$. Even for small $L = 24$, a clear negative peak can be developed, indicating a strong first-order transition behavior. Furthermore, for a fixed $L$, the peak of $U_d$ grows with increasing $J_3$, as clearly shown in Fig. 10(b).

On the other hand, for the case of $J_3 > 0.5$, the double stripe state and the plaquette state are degenerated in the $J_1$-$J_2$-$J_3$ model on the square lattice. To study the transition behaviors of the double stripe state or the plaquette state, the degeneracy of these two states should be broken. In some extent, some other frustrated spin models such as the well-known Shastry-Sutherland model may be studied to investigate the phase transitions.[33] However, this topic is beyond the scope of this work, and will be left for our future work.

## IV. Conclusion

In Conclusion, the role of the third nearest neighbor interaction on the phase transitions and critical behaviors of the frustrated $J_1$-$J_2$-$J_3$ model on the square lattice is investigated using Monte Carlo simulations. In a certain range of $J_3$, the critical exponents of the continuous transition vary with the increase of the transition temperature, exhibiting an Ashkin-Teller-like behavior. In addition, the transition points at the Potts-critical end point estimated based on the analytic method are very similar for every $J_2$. Thus, this work suggests that the critical behaviors may be closely dependent on the transition point for a fixed $J_1$. Furthermore, a strong first-order behavior is confirmed for the transition to the staggered dimer state which is stabilized by an antiferromagnetic $J_3$.

**Acknowledgements**:

This work was supported by the Natural Science Foundation of China (51332007, 51322206, 11274094), and the National Key Projects for Basic Research of China (2015CB921202 and 2015CB654602).

*References:*

1  D. Friedan, Z. Qiu, and S. Shenker, Phys. Rev. Lett. 52, 1575 (1984).
2  H. T. Diep, Frustrated spin systems (World Scientific, Singapore) 2004.
3  D. P. Landau and K. Binder, *A Guide to Monte Carlo Simulations in Statistical Physics* (Cambridge University Press, Cambridge, England, 2008).
4  H. Y. Wang, Phys. Rev. B 86, 144411 (2012).
5  M. P. Nightingale, Phys. Lett. A 59, 486 (1977).
6  S. B. Jin, A. Sen, and A. W. Sandvik, Phys. Rev. Lett. 108, 045702 (2012).
7  R. H. Swendsen and S. Krinsky, Phys. Rev. Lett. 43, 177 (1979).
8  K. Binder and D. P. Landau, Phys. Rev. B 21, 1941 (1980).
9  J. Oitmaa, J. Phys. A 14, 1159 (1981).
10  D. P. Landau and K. Binder, Phys. Rev. B 31, 5946 (1985).
11  J. L. Morán-López, F. Aguilera-Granja, and J. M. Sanchez, Phys. Rev. B 48, 3519 (1993).
12  A. Malakis, P. Kalozoumis, and N. Tyraskis, Eur. Phys. J. B 50, 63 (2006).
13  A. Kalz, A. Honecker, S. Fuchs, and T. Pruschke, Eur. Phys. J. B 65, 533 (2008).
14  A. Kalz, A. Honecker, and M. Moliner, Phys. Rev. B 84, 174407 (2011).
15  J. Ashkin and E. Teller, Phys. Rev. 64, 178 (1943).
16  S. B. Jin, A. Sen, W. N. Guo, and A. W. Sandvik, Phys. Rev. B 87, 144406 (2013).
17  R. J. Baxter, J. Phys. C 6, 445 (1973).
18  A. Kalz and A. Honecker, Phys. Rev. B 86, 134410 (2012).
19  R. Yu and Q. M. Si, Phys. Rev. Lett. 115, 116401 (2015).
20  F. J. Ma, W. Ji, J. P. Hu, Z. Y. Lu, and T. Xiang, Phys. Rev. Lett. 102, 177003 (2009).
21  N. Metropolis, A. W. Rosenbluth, M. N. Rosenbluth, A. H. Teller, and E. Teller, J. Chem. Phys. 21, 1087 (1953).
22  K. Hukushima and K. Nemoto, J. Phys. Soc. Jpn. 65, 1604 (1996).
23  L. Huo, W. C. Huang, Z. B. Yan, X. T. Jia, X. S. Gao, M. H. Qin and J.-M. Liu, J. Appl. Phys. 113, 073908 (2013).
24  R. J. Baxter, *Exactly Solved Models in Statistical Mechanics* (Academic Press, London,

1982).

25  S. Wiseman and E. Domany, Phys. Rev. E 48, 4080 (1993).

26  K. Vollmayr, J. D. Reger, M. Scheucher, and K. Binder, Z. Phys. B 91, 113 (1993).

27  A. L. Wysocki, K. D. Belashchenko, and V. P. Antropov, Nat. Phys. 7, 485 (2011).

28  E. Dagotto, Rev. Mod. Phys. 85, 849 (2013).

29  M. H. Qin, S. Dong, H. B. Zhao, Y. Wang, J.-M. Liu, and Z. F. Ren, New J. Phys. 16, 053027 (2014).

30  P. C. Dai, Rev. Mod. Phys. 87, 855 (2015).

31  J. K. Glasbrenner, I. I. Mazin, H. O. Jeschke, P. J. Hirschfeld, R. M. Fernandes, and R. Valenti, Nat. Phys. 11, 953 (2015).

32  F. A. Kassan-Ogly, A. K. Murtazaev, A. K. Zhuravlev, M. K. Ramazanov, and A. I. Proshkin, J. Magn. Magn. Mater. 384, 247 (2015).

33  B. S. Shastry and B. Sutherland, Physica B & C 108, 1069 (1981).

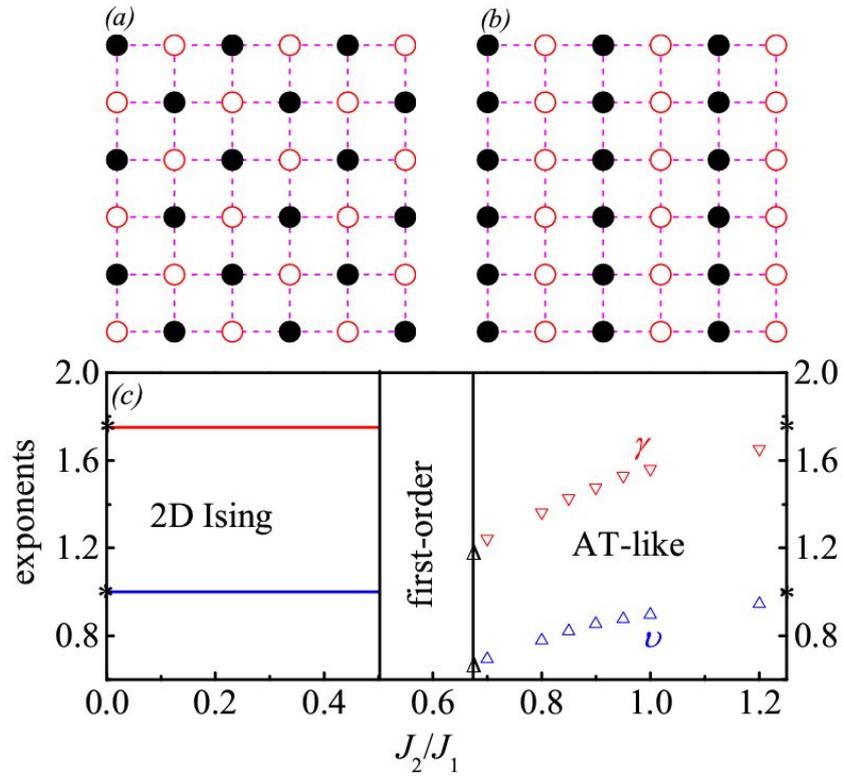

Fig.1. (color online) Spin configurations in the (a) Néel state, and (b) single stripe state. Solid and empty circles represent the up-spins and the down-spins, respectively. (c) The critical behaviors of the frustrated $J_1$-$J_2$ model on the square lattice. The Potts values (black triangles) and the Ising values (black stars) are also given, and these critical exponents for $J_2/J_1 > 0.67$ are reproduced from Ref. 18.

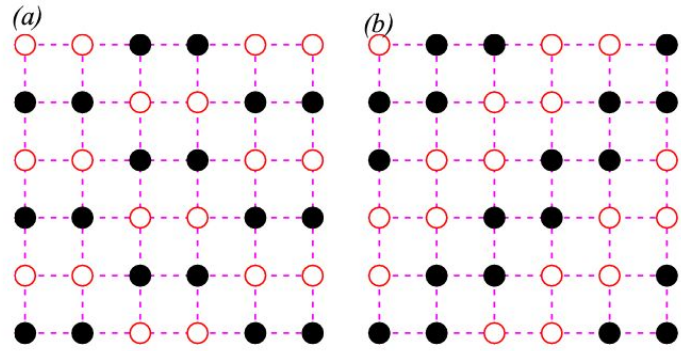

Fig.2. (color online) Spin configurations in the (a) staggered dimer state, and (b) double stripe state. Solid and empty circles represent the up-spins and the down-spins, respectively.

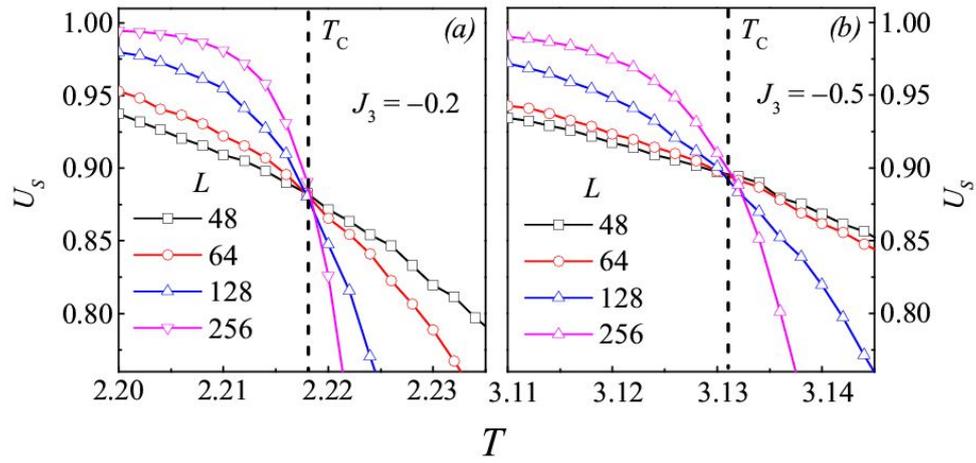

Fig.3. (color online) Binder cumulant $U_s$ as a function of $T$ for different $L$ at $J_2 = 0.8$ at (a) $J_3 = -0.2$ and (b) $J_3 = -0.5$.

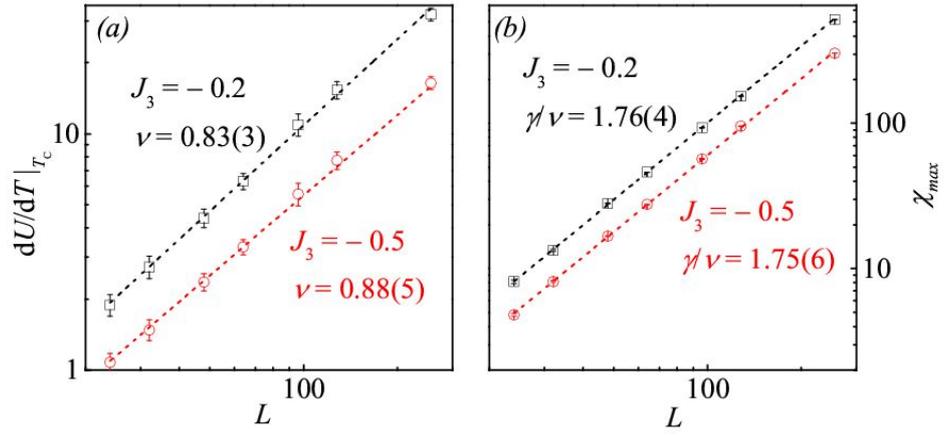

Fig.4. (color online) Log-log plot of (a) d$U$/d$T$($T_C$), and (b) $\chi_{max}$ for various $L$ at $J_3$ = 0.2 and $J_3$ = 0.5 for $J_2$ = 0.8.

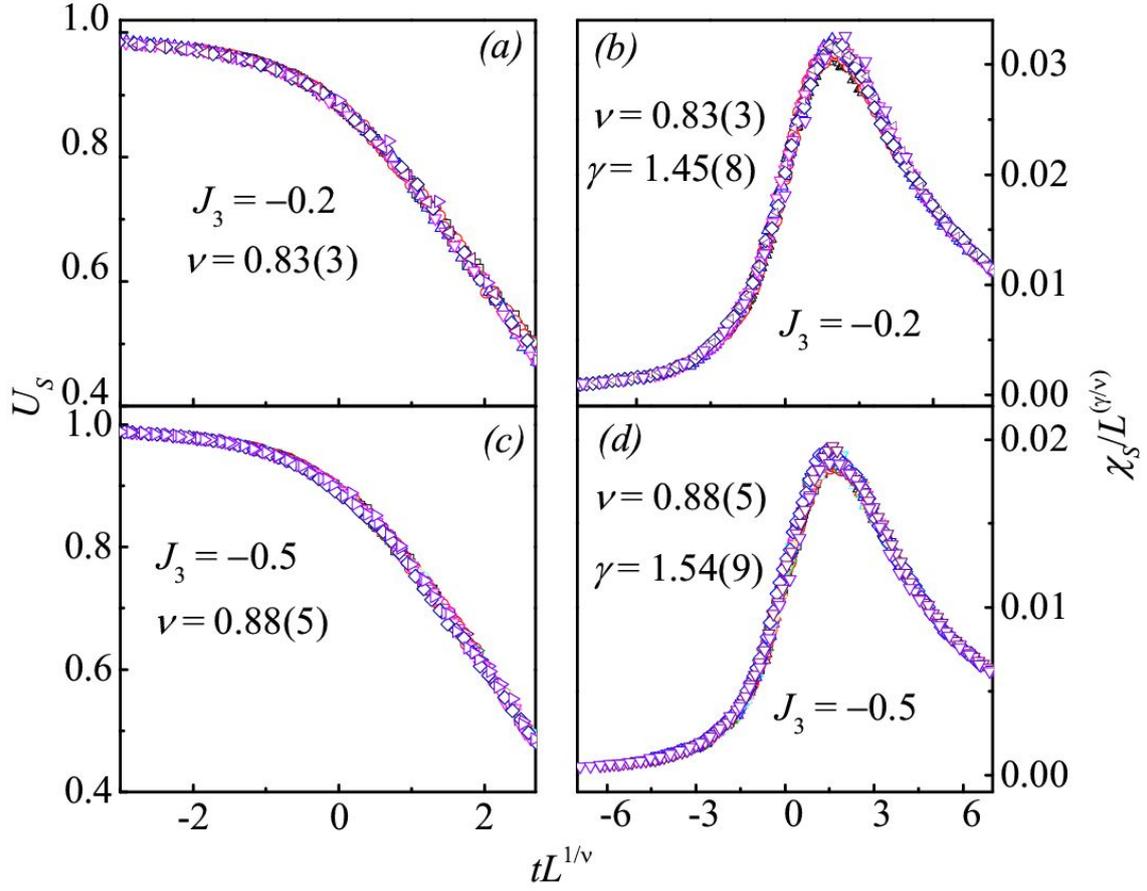

Fig.5. (color online) A scaling plot of $U_s$ ((a) and (c)), and $\chi_s$ ((b) and (d)) at $J_3 = -0.2$ ((a) and (b)) and $J_3 = -0.5$ ((c) and (d)) at $J_2 = 0.8$.

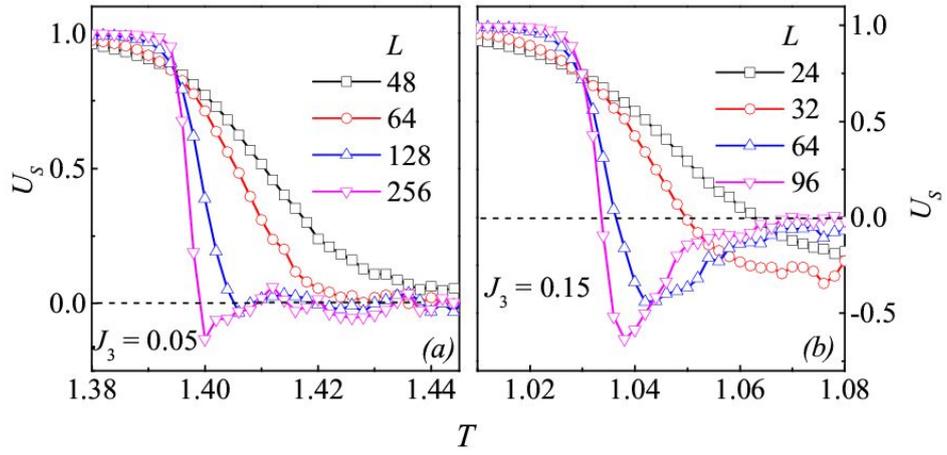

Fig.6. (color online) Binder cumulant $U_s$ as a function of $T$ for different $L$ at $J_2 = 0.8$ at (a) $J_3 = 0.05$ and (b) $J_3 = 0.15$.

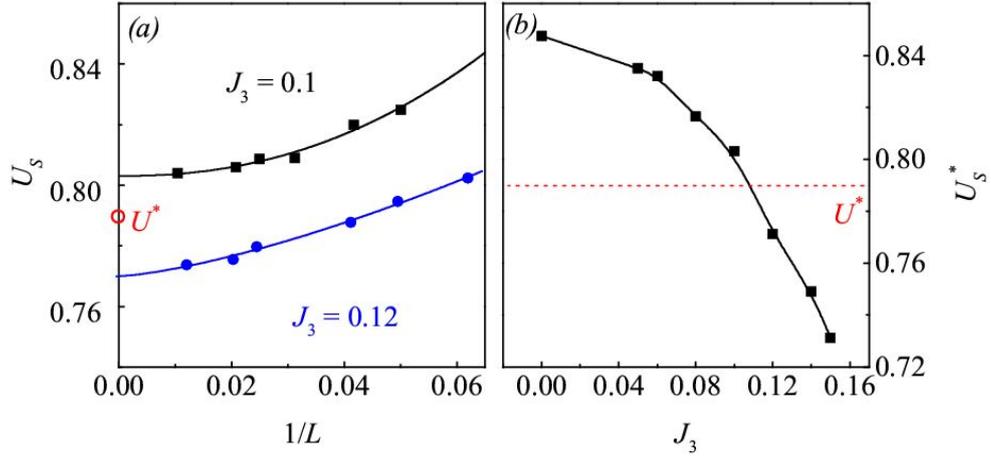

Fig.7. (color online) (a) Binder cumulant crossing points $U_s$ for ($L$, $2L$) system pairs and the extrapolation to $L = \infty$, and (b) $U_s^*$ of the $J_1$-$J_2$-$J_3$ model for various $J_3$ at $J_2 = 0.8$ compared with that of the 4-state Potts model.

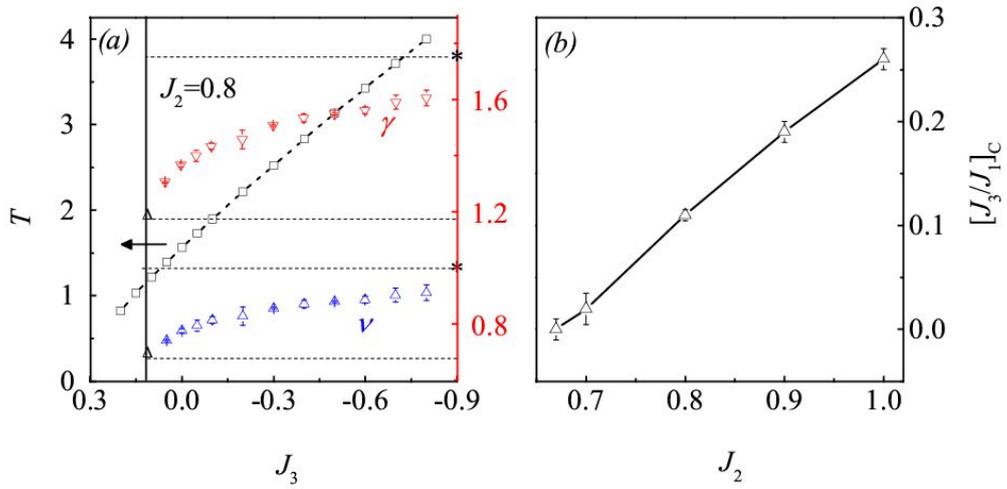

Fig.8. (color online) (a) Phase diagram in the ($J_3$, $T$) parameter plane at $J_2 = 0.8$. The critical exponents $v$ and $\gamma$ of the $J_1$-$J_2$-$J_3$ model, 4-state potts model (black triangles) and Ising model (black stars) are also given. (b) The estimated Potts-critical end points $[J_3/J_1]_C$ for various $J_2$.

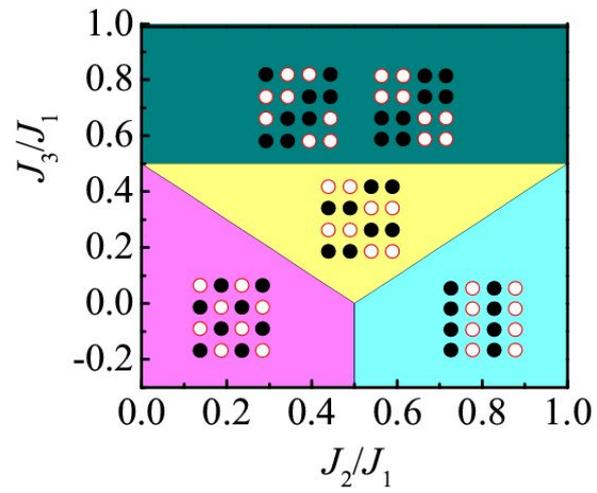

Fig.9. (color online) Ground-state phase diagram in the ($J_2$, $J_3$) parameter plane. The spin configurations of these states are depicted. Solid and empty circles represent the up-spins and the down-spins, respectively.

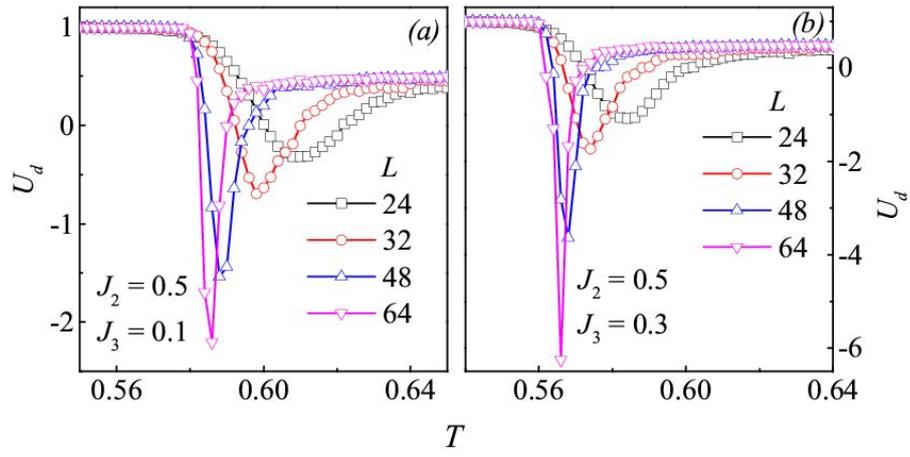

Fig.10. (color online) Binder cumulant $U_d$ as a function of $T$ for different $L$ at $J_2 = 0.5$ at (a) $J_3 = 0.1$ and (b) $J_3 = 0.3$.